\documentclass[sigconf]{acmart}
\usepackage{enumitem}
\AtBeginDocument{%
  }

\setcopyright{acmlicensed}
\copyrightyear{2018}
\acmYear{2018}
\acmDOI{XXXXXXX.XXXXXXX}
\acmConference[Conference acronym 'XX]{Make sure to enter the correct
  conference title from your rights confirmation email}{June 03--05,
  2018}{Woodstock, NY}
\acmISBN{978-1-4503-XXXX-X/2018/06}

\begin{document}

\title[AgentEconomist: End-to-End Economic Intuition Translation]{AgentEconomist: An End-to-end Agentic System Translating Economic Intuitions into Executable Computational Experiments}

\author{Jiaju Chen}
\authornote{Equal contribution.}
\email{cjj01@mail.ustc.edu.cn}
\affiliation{%
  \institution{Zhongguancun Academy}
  \city{Beijing}
  \country{China}}
\affiliation{%
  \institution{University of Science and Technology of China}
  \city{Hefei}
  \country{China}}

\author{Jinghua Piao}
\authornotemark[1]
\email{pjh22@mails.tsinghua.edu.cn}
\affiliation{%
  \institution{Zhongguancun Academy}
  \city{Beijing}
  \country{China}}
\affiliation{%
  \institution{Department of Electronic Engineering, Tsinghua University}
  \city{Beijing}
  \country{China}}

\author{Xia Xu}
\email{s-xx25@bza.edu.cn}
\affiliation{%
  \institution{Zhongguancun Academy}
  \city{Beijing}
  \country{China}}

\author{Songwei Li}
\email{lisw21@mails.tsinghua.edu.cn}
\affiliation{%
  \institution{Zhongguancun Academy}
  \city{Beijing}
  \country{China}}
\affiliation{%
  \institution{Department of Electronic Engineering, Tsinghua University}
  \city{Beijing}
  \country{China}}

\author{Tong Xia}
\email{Tongxia@tsinghua.edu.cn}
\affiliation{%
  \institution{Zhongguancun Academy}
  \city{Beijing}
  \country{China}}
\affiliation{%
  \institution{Vanke School of Public Health, Tsinghua University}
  \city{Beijing}
  \country{China}}

\author{Xiangnan He}
\authornote{Corresponding authors.}
\email{xiangnanhe@gmail.com}
\affiliation{%
  \institution{Zhongguancun Academy}
  \city{Beijing}
  \country{China}}
\affiliation{%
  \institution{University of Science and Technology of China}
  \city{Hefei}
  \country{China}}

\author{Yong Li}
\authornotemark[2]
\email{liyong07@tsinghua.edu.cn}
\affiliation{%
  \institution{Zhongguancun Academy}
  \city{Beijing}
  \country{China}}
\affiliation{%
  \institution{Department of Electronic Engineering, Tsinghua University}
  \city{Beijing}
  \country{China}}

\renewcommand{\shortauthors}{Chen et al.}

\begin{abstract}
A long-standing challenge in economics lies not in the lack of intuition, but in the difficulty of translating intuitive insights into verifiable research. To address this challenge, we introduce \textbf{AgentEconomist}, an end-to-end interactive system designed to translate abstract intuitions into executable computational experiments. Grounded in a domain-specific knowledge base covering over 13,000 high-quality academic papers, the system employs a modular multi-stage architecture. Specifically, the Idea Development Stage generates literature-grounded hypotheses, the Experimental Design Stage configures simulator-aligned experimental parameters and protocols, and the Experimental Execution Stage runs experiments and returns structured analyses. Together, these stages form a human-in-the-loop, iterative workflow that translates economic intuitions into executable computational experiments. Through extensive experiments involving human expert evaluation and large language models (LLMs) as judges, we show that the system generates research ideas with stronger literature grounding and higher novelty and insight than state-of-the-art generic LLMs. Overall, AgentEconomist adopts a human-AI collaboration paradigm that enables researchers to focus on high-level intuitions, while delegating the labor-intensive processes of translation and computational execution to agents.
\end{abstract}



\received{20 February 2007}
\received[revised]{12 March 2009}
\received[accepted]{5 June 2009}

\maketitle

\section{Introduction}

\begin{figure}[t]
    \centering
    \includegraphics[width=0.5\textwidth]{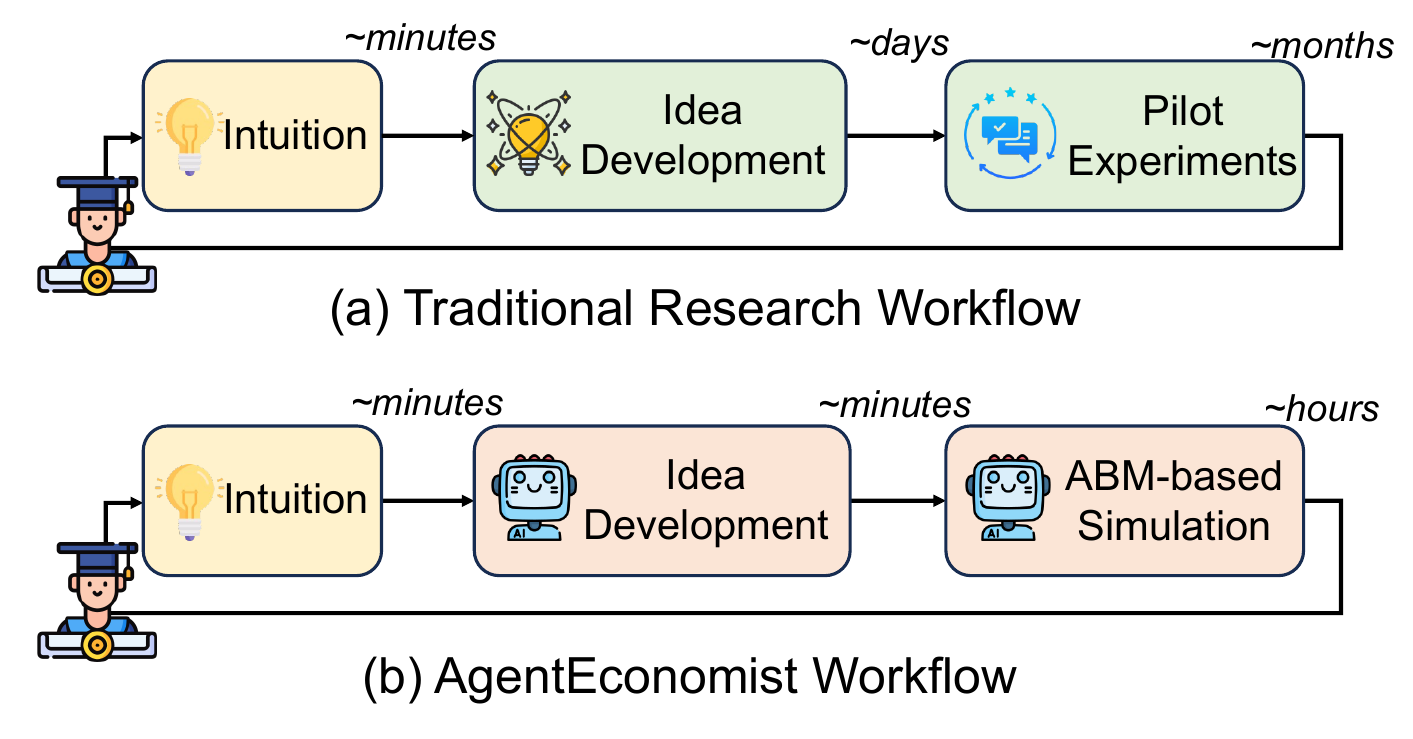} 
    \vspace{-5mm}
    \caption{
    Comparison of economic research workflows.
    ({a}) In traditional workflows, translating intuition into pilot experiments involves slow and resource-intensive idea development and experimentation cycles.
    ({b}) AgentEconomist shortens the intuition-to-simulation loop by supporting rapid idea development and agent-based simulation, enabling faster hypothesis verification.
    }
    \label{fig:workflow_comparison} 
    \vspace{-5mm}
\end{figure}

Scientific research often begins with a spark of intuition, a pre-formal, qualitative understanding of how underlying mechanisms in the world function, which serves as the starting point of any formal ideas and hypotheses~\citep{polanyi2009tacit,popper2005logic}. In economics, such intuitions typically concern how agents respond to incentives~\citep{becker1976economic}, how institutions shape behavior~\citep{north1990institutions}, and how these interactions give rise to macro-level outcomes~\citep{schelling2006micromotives}. To move beyond mere conjecture, economists traditionally rely on theoretical derivation~\citep{samuelson1948foundations} or field experiments~\cite{harrison2004field,duflo2011poor}. However, as illustrated in Figure~\ref{fig:workflow_comparison}, this traditional workflow is inherently slow and resource-intensive, thereby limiting the scope of ideas a researcher can explore. Computational experiments, based on agent-based modeling (ABM), are a well-established methodology in economics and finance for simulation-based analysis~\citep{tesfatsion2006agent,lebaron2006agent,farmer2009economy} and are frequently employed in teaching to convey complex economic mechanisms~\citep{tisue2004netlogo,epstein1996growing,railsback2019agent}. Yet, while such simulation environments provide a powerful medium for verifying economic insights, their inherent complexity often hinders their practical feasibility in research workflows~\citep{axtell2025agent}.

This barrier between intuition and computational experiments manifests in three distinct dimensions in traditional modeling workflows. First, the process of idea development is often implicit. Novices struggle to systematize their intuition into concrete modeling choices because key methodological knowledge—such as selecting plausible mechanisms or assumptions—remains ``tacit knowledge'' locked in the minds of experts, rarely explicit in textbooks. Second, turning an idea into a computational experiment is itself challenging. Even when a researcher possesses a clear idea, translating it into an executable simulation experiments involves navigating complex codebases, creating a friction that slows down hypothesis testing. Third, and perhaps most critically, the research process lacks systematic experience accumulation. Scientific discovery is an iterative path of refinement, yet the rationale behind specific iterations, why a parameter was tweaked or a hypothesis rejected, is poorly organized alongside the simulation experiments. This loss of \textit{epistemic context} forces researchers to rely on fragmented memory rather than a structured history of inquiry, making it difficult to learn from past failures or incrementally refine ideas.

Recent advances in leveraging LLM agents for autonomous knowledge discovery have attempted to streamline scientific workflows, yet they reveal substantial limitations when assisting researchers in such complex domains. First, existing efforts often target isolated stages, e.g., optimizing either hypothesis generation~\citep{gottweis2025towards} or code execution~\citep{swanson2025virtual}, rather than supporting the seamless workflow required to link literature to simulation. Second, holistic frameworks like \textit{The AI Scientist}~\citep{lu2024ai} tend to prioritize ``outcome automation'' (e.g., generating a final manuscript) over ``process support''. By treating research as a black box to be automated, they neglect the critical need for \textit{sense-making}: helping researchers structure their thinking and understand the ``why'' behind modeling decisions. Finally, current systems largely pursue generic applicability at the expense of domain depth, lacking the specific grounding required for economics, where research is deeply rooted in theoretical frameworks and institutional constraints.

To address these challenges, we focus on economics as a representative domain and introduce \textbf{AgentEconomist}, an end-to-end interactive research copilot for economic simulation.
Built on top of AgentEconomy, a comprehensive agent-based economic simulator, AgentEconomist is designed to support the entire intuition-to-experiment workflow rather than replacing the researcher.
Crucially, the system grounds idea development in a large-scale corpus of economic literature, leveraging over 13,000 academic papers from top-tier journals to make tacit theoretical knowledge explicit and accessible.
The workflow is decomposed into three specialized stages.
The \emph{Idea Development Stage} supports literature-grounded sense-making by retrieving relevant economic studies and synthesizing mechanisms, assumptions, and variables.
The \emph{Experimental Design Stage} formalizes these ideas into testable hypotheses and executable experiment specifications.
The \emph{Experimental Execution Stage} operationalizes the design by running simulations and returning structured results.
To support continuity and reliable execution, we introduce a \emph{Structured Memory} module that preserves theoretical context, experimental decisions, and outcomes across iterations, and an \emph{MCP-based toolbox} that standardizes simulator interaction for configuration, execution, and result collection.

Evaluating such open-ended research assistants presents a unique challenge, as standard benchmarks cannot capture the nuance of scientific reasoning. Therefore, we design a rigorous \textbf{mixed-methods evaluation protocol}. First, to assess the quality of Idea Generation, we employ a dual-evaluation framework where both advanced LLMs and human experts score generated hypotheses across eight distinct dimensions (e.g., economic soundness, novelty, and feasibility). Second, to validate the system's utility in real-world workflows, we conduct a holistic user study. Through questionnaires and semi-structured interviews, we gather qualitative feedback on the researchers' experience, focusing on how the system affects their cognitive load, sense of agency, and overall research efficiency. Third, to validate end-to-end scientific usefulness, we report a real-session case study showing that the system can generate interpretable, mechanism-consistent findings from user-posed intuitions. Across these protocols, our results show that AgentEconomist consistently outperforms strong baselines on the core dimensions of hypothesis novelty and literature grounding, while also producing hypotheses that are more readily operationalizable in simulation-based settings. We have released our code.\footnote{\url{https://github.com/Jiaju-Chen/AgentEconomist}}
The primary contributions of this work include:
\begin{itemize}[leftmargin=8pt]
\item \textbf{A human-AI collaboration workflow for bridging the intuition-execution gap.} We conceptualize the translation of economic intuition into computational experiments as a collaborative process. By explicitly modeling the transition from implicit sense-making to operational execution, we provide a structured workflow that lowers the technical barriers for economic modeling.

\item \textbf{An end-to-end system architecture for grounded experimentation.} We introduce \textit{AgentEconomist}, a unified system that integrates a retrieval-augmented knowledge base with specialized agents. This design ensures that generated hypotheses are theoretically sound and that designed experiments are executable, forming a closed loop for iterative discovery.

\item \textbf{Empirical validation via mixed-methods assessment and case study evidence.} We conduct quantitative evaluations across 8 quality dimensions, qualitative user interviews, and an end-to-end real-session case study, demonstrating that AgentEconomist effectively produces higher-quality research ideas, reveals interpretable scientific phenomena, and significantly streamlines the research workflow.
\end{itemize}

\section{Related Works}

\subsection{Task-Specific Scientific Assistants}
Recent research has focused on augmenting specific stages of the research lifecycle. 
In the realm of data management and deep research \citep{huang2025deep}, specific systems like \textit{ChatPD} \citep{xu2025chatpd} and \textit{SciSciGPT} \citep{shao2025sciscigpt} have emerged to automate dataset discovery and literature analysis, though benchmarks like \textit{DATASETRESEARCH} \citep{li2025datasetresearch} reveal that current agents still struggle significantly with out-of-distribution dataset demands.
For ideation, methods have evolved from iterative prompting \citep{gottweis2025towards} to multi-agent frameworks like \textit{VIRSCI} \citep{su2025many}, which simulate team collaboration to enhance idea novelty.
In the evaluation phase, approaches vary by technical architecture: \textit{ReviewRL} \citep{zeng2025reviewrl} employs reinforcement learning to optimize feedback quality, while \textit{ReviewAgents} \citep{gao2025reviewagents} and \textit{DeepReview} \citep{zhu2025deepreview} utilize multi-agent collaboration and chain-of-thought reasoning to bridge the gap with human peer reviews.
{Crucially}, however, assessments like \textit{IdeaBench} \citep{guo2025ideabench} expose a paradox: while LLMs generate highly novel ideas, they often lack practical feasibility. 
This highlights a limitation of such ``open-loop'' tools: they produce concepts without the agency to validate them. 
{In contrast}, AgentEconomist bridges this gap by coupling literature-grounded ideation with an execution toolbox, ensuring abstract intuitions are operationalized into testable simulations.
More specifically, unlike general-purpose assistants that rely on broad web-scale retrieval or generic priors, our framework grounds hypothesis formulation and experiment design in a curated, indexed economics corpus.

\begin{figure*}[t]
\centering
\includegraphics[width=\textwidth]{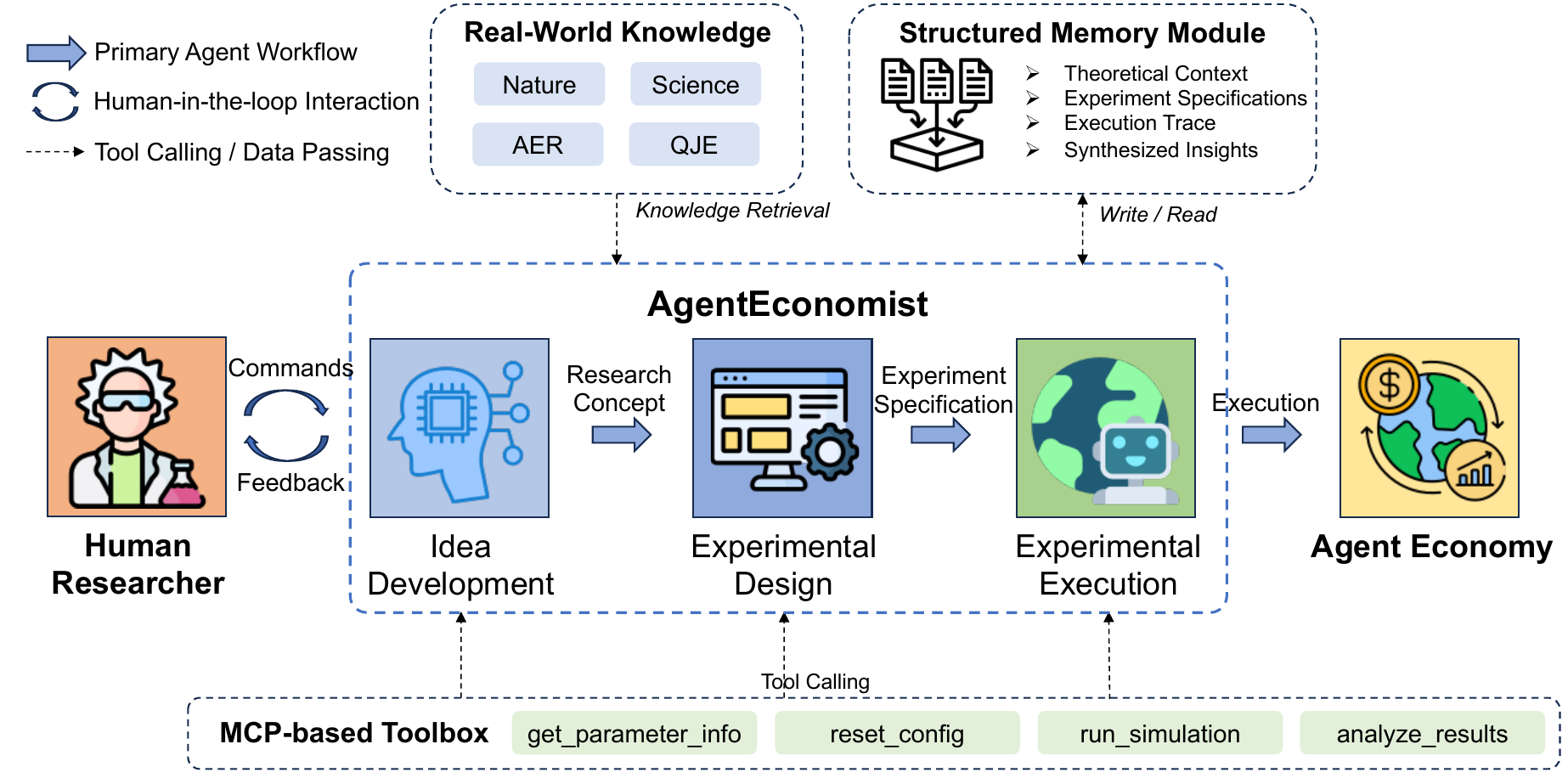}
\caption{
Overview of the AgentEconomist framework.
The system structures the intuition-to-experiment process into three stages: literature-grounded idea development, experimental design formalization, and experimental execution.
These stages are handled by specialized designs that coordinate through a structured memory module and interact with the simulator via an MCP-based toolbox, while keeping the human researcher in the loop.
}
\label{fig:framework}
\vspace{6mm}
\end{figure*}

\subsection{Autonomous Research Systems}
A parallel stream of work aims to replicate the full role of a human researcher through end-to-end automation \citep{hu2025survey}. 
Building on the code-generation capabilities of \textit{The AI Scientist} \citep{lu2024ai}, recent systems have achieved significant breakthroughs: \textit{CycleResearcher} \citep{weng2024cycleresearcher} introduced a closed-loop ``research-review-revise'' mechanism, while \textit{DeepScientist} \citep{weng2025deepscientist} incorporated Bayesian optimization for long-term discovery. 
To support these autonomous agents, infrastructure like \textit{ToolUniverse} \citep{gao2025democratizing} has emerged to standardize tool usage. 
Furthermore, recognizing the complexity of scientific inquiry, frameworks like \textit{OmniScientist} \citep{shao2025omniscientist} and \textit{MirrorMind} \citep{zeng2025mirrormind} have expanded this vision by integrating diverse agents to model professional research workflows and collaborative reasoning.
{However}, critical gaps remain when applying these generalist systems to economic inquiry. 
Existing models often function as black boxes that prioritize outcome automation over process support, denying novices the opportunity for sense-making. 
Moreover, they lack the deep domain grounding required to map theoretical constructs to high-dimensional simulation parameters. 
{To address this}, AgentEconomist positions itself as an interactive co-pilot grounded in a specialized knowledge base of over 13k academic papers. 
By coupling domain depth with rigorous execution, it ensures that scientific discovery remains a transparent, human-aligned process.
Crucially, our primary differentiator is tight coupling to an executable simulation substrate: the system maps hypotheses to concrete simulator parameters, checks feasibility constraints before plan finalization, executes controlled runs, and returns comparable outcomes rather than stopping at ideation or draft plans.
In addition, we maintain a manifest-based structured memory that records retrieved paper IDs, parameter choices, configuration artifacts, and execution outputs, enabling traceable multi-round refinement beyond unstructured conversational context.

\section{Method}
\label{sec:method}
\subsection{Design Rationale: Human--Agent Complementarity}
\label{sec:rationale}

Economic research involves heterogeneous cognitive demands.
Human researchers excel at forming high-level intuitions, exercising normative judgment, and deciding when an explanation or result is sufficient.
In contrast, transforming these intuitions into executable experiments requires systematic literature grounding, formal specification, and reliable execution—tasks that are labor-intensive and error-prone when performed manually.

AgentEconomist is designed around this complementarity.
The system assigns abstraction, judgment, and goal-setting to the human researcher, while delegating literature-grounded sense-making, experiment formalization, and execution management to automated components.
This separation allows each part of the workflow to be handled by the entity best suited for it, while preserving human control over research direction and interpretation.

\subsection{The AgentEconomist Framework}
\label{sec:framework}

Building on the design rationale above, we introduce the AgentEconomist framework shown in Figure~\ref{fig:framework}.
The framework operationalizes the intuition-to-experiment workflow through a modular multi-stage architecture, supported by a shared infrastructure layer.

\subsubsection{Agent Core}

The cognitive workflow is decomposed into three specialized stages corresponding to key phases of economic research:

\begin{itemize}[leftmargin=8pt]
    \item \textbf{Idea Development Stage.}
    This stage starts from a user-provided coarse intuition and performs literature-grounded sense-making with feasibility awareness.
    It retrieves relevant evidence via retrieval-augmented generation (RAG), proposes candidate hypotheses, and checks executable simulation parameters from AgentEconomy to assess operationalizability.
    To enforce methodological validity, hypothesis generation is constrained by a capability boundary: all claims must be expressible through simulator-supported variables and interventions.
    We adopt a parameter-first strategy in this stage, requiring the agent to map intuition into implementable parameter candidates before forming causal hypothesis statements.
    The stage then yields either a refined, testable hypothesis with mechanism-consistent expected outcomes, or a structured feasibility diagnosis that highlights violated constraints and guides user revision.
    If no executable hypothesis can be produced under current simulator constraints, the agent must explicitly report why generation fails (e.g., missing variables, unsupported interventions, or inconsistent assumptions) and provide a minimal, clearly labeled proxy direction for user revision.

    \item \textbf{Experimental Design Stage.}
    Building on the validated hypothesis, this stage formalizes the study into simulator-ready specifications.
    It maps hypotheses to concrete configurations, defines experiment structures (e.g., control/treatment groups, multi-condition settings, and hyperparameter sweeps), and aligns evaluation metrics with simulation semantics.
    To preserve identifiability, each experimental condition is constrained to modify only the minimum necessary variables, reducing multi-mechanism confounding across groups.
    The stage also enforces metric alignment: each hypothesized dependent variable must be mapped to explicitly computable simulator metrics rather than qualitative or generic statements.
    The resulting design specifies parameter settings, group structure, run plans, and metric definitions for downstream execution.

    \item \textbf{Experimental Execution Stage.}
    Given the formalized design, this stage operationalizes experiments through the MCP-based toolbox over AgentEconomy.
    It handles configuration, execution-state management, and run-time trace collection, and returns structured logs together with metric-aligned visualized outputs.
    For reproducibility, executions use fixed random seeds, and each result artifact is bound to its versioned configuration.
    For traceability, the stage records input configurations, execution states, error categories, and result summaries as structured logs.
    To prevent over-interpretation, reporting is restricted to pre-registered metrics and does not extend to subjective policy narratives.
    These artifacts support direct interpretation and iterative follow-up in subsequent research cycles.
\end{itemize}

\subsubsection{Agent Infrastructure}

The agent core is supported by three foundational infrastructure components that ensure grounding, continuity, and reliable execution:

\begin{itemize}[leftmargin=8pt]
    \item \textbf{Real-World Knowledge Base.}
    To support literature-grounded reasoning, we construct a domain-specific RAG pipeline over a vector database containing over 13,000 academic papers from top-tier economics and interdisciplinary journals.
    For each paper, we chunk the abstract and introduction, encode chunks with citation-informed SPECTER2 embeddings\cite{Singh2022SciRepEvalAM} (768 dimensions), and index them using cosine similarity retrieval.
    We further apply metadata filtering (e.g., publication year and venue) and maintain manifest logs of retrieved paper IDs to ensure traceability and reproducibility.
    Retrieved evidence supports both idea development and experimental design by making relevant mechanisms, assumptions, and empirical precedents explicit.

    \item \textbf{Structured Memory Module.}
    Unlike embedding-centric memory systems that primarily store interaction history as dense vectors for semantic recall (e.g., Mem0~\citep{chhikara2025mem0}), our setting requires precise localization of assumptions, parameter constraints, and decision rationales.
    Therefore, we maintain a structured textual memory that explicitly records theoretical context, experiment specifications, execution traces, and synthesized outcomes.
    This design supports context-preserving iteration and cumulative reasoning across research cycles while allowing agents to directly reference exact prior decisions instead of approximate nearest-neighbor recalls.

    \item \textbf{MCP-based Toolbox.}
    AgentEconomist interacts with the simulator through a standardized toolbox built on the Model Context Protocol (MCP), which provides a stable interface between language agents and executable simulation tools.
    The toolbox abstracts low-level simulator APIs into high-level semantic actions, including parameter inspection, environment initialization, experiment configuration, job execution, status polling, log collection, and result export.
    This abstraction reduces implementation complexity for agent planning, enforces consistent invocation patterns across iterations, and improves reliability by making execution states observable and recoverable.
    In addition, by standardizing tool calls and returned artifacts, the MCP layer supports reproducible reruns and simplifies downstream analysis and auditing.

\end{itemize}

\begin{figure*}[!t]
    \centering
    \includegraphics[width=\textwidth]{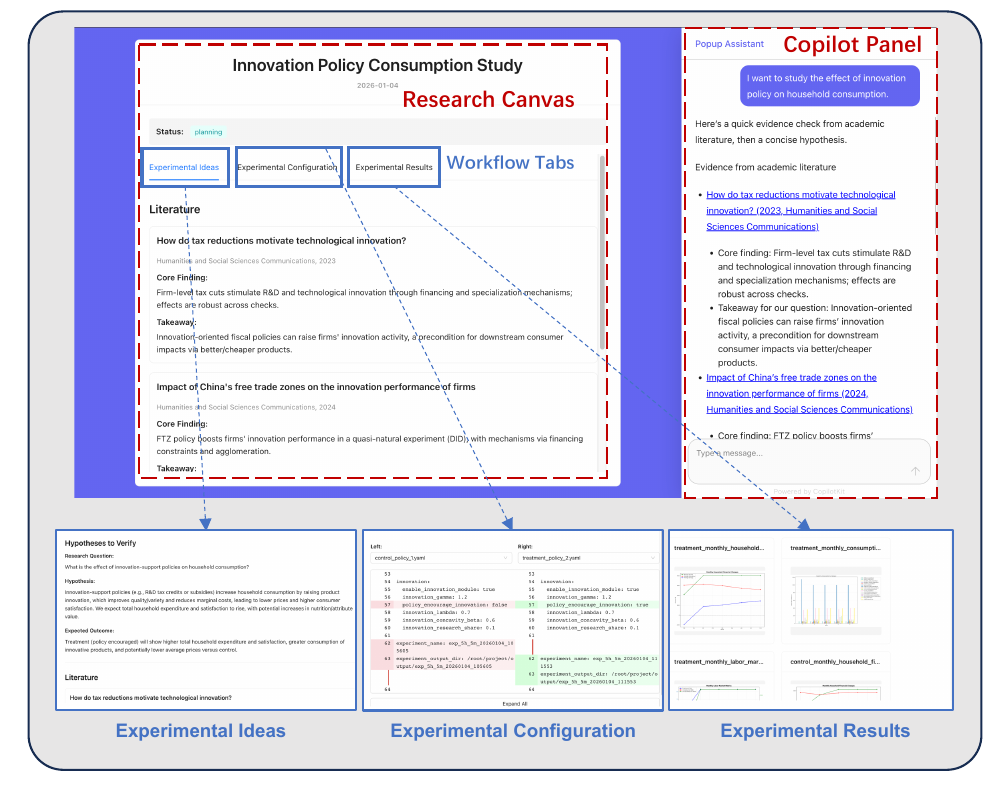}
    \caption{Detailed front-end component view. The interface combines a research canvas with workflow tabs, a copilot panel for human--agent interaction, and dedicated views for experimental ideas, configuration, and results.}
    \label{fig:frontend_display}
\vspace{6mm}
\end{figure*}

\subsection{Simulation Substrate: AgentEconomy}
\label{sec:simulation}

To support the execution and evaluation of experiments specified by AgentEconomist, we develop AgentEconomy as a computational laboratory for rapid hypothesis verification.
Unlike toy models that isolate specific sectors, AgentEconomy provides a comprehensive and flexible agent-based environment designed to capture the ripple effects of economic interventions across interacting components.

\paragraph{Comprehensive Economic Ecosystem.}
The environment models a closed-loop economic system comprising four core entity types: households, firms, a government, and a bank.
These entities interact through two explicitly modeled markets grounded in real-world data:
\begin{itemize}[leftmargin=8pt]
    \item \textbf{Labor Market:} Matches household skills to firm job requirements, allowing wages and employment to emerge endogenously from supply--demand dynamics.
    \item \textbf{Product Market:} Facilitates the exchange of goods, where consumption is driven by heterogeneous preferences rather than stylized aggregate functions.
\end{itemize}
This comprehensive coverage enables AgentEconomist to test hypotheses involving multi-stage transmission channels, such as how policy interventions affect household outcomes through firm-level and labor-market adjustments.

\paragraph{Flexible LLM-Driven Behavior.}
Crucially, agents in AgentEconomy do not rely on rigid, hard-coded heuristics.
Instead, they operate through a hybrid mechanism that combines LLM-based reasoning with empirical grounding.
Initialized with microdata (e.g., PSID profiles) to ensure realism, agents reason adaptively about decisions such as labor supply and pricing in response to changing incentives.
This flexibility allows the system to simulate plausible responses to novel policy shocks (e.g., AI regulation, universal basic income) that are difficult to encode using fixed behavioral rules.

\paragraph{Testbed for Rapid Experimentation.}
By combining structural completeness with behavioral flexibility, AgentEconomy functions as a high-fidelity testbed for hypothesis-driven experimentation.
It exposes a rich set of tunable parameters (policy levers) and observable metrics (macro-level outcomes), enabling AgentEconomist to translate abstract intuitions into concrete simulation runs and receive immediate feedback on validity and emergent phenomena.

\begin{figure*}[!t]
    \centering
    \includegraphics[width=\textwidth]{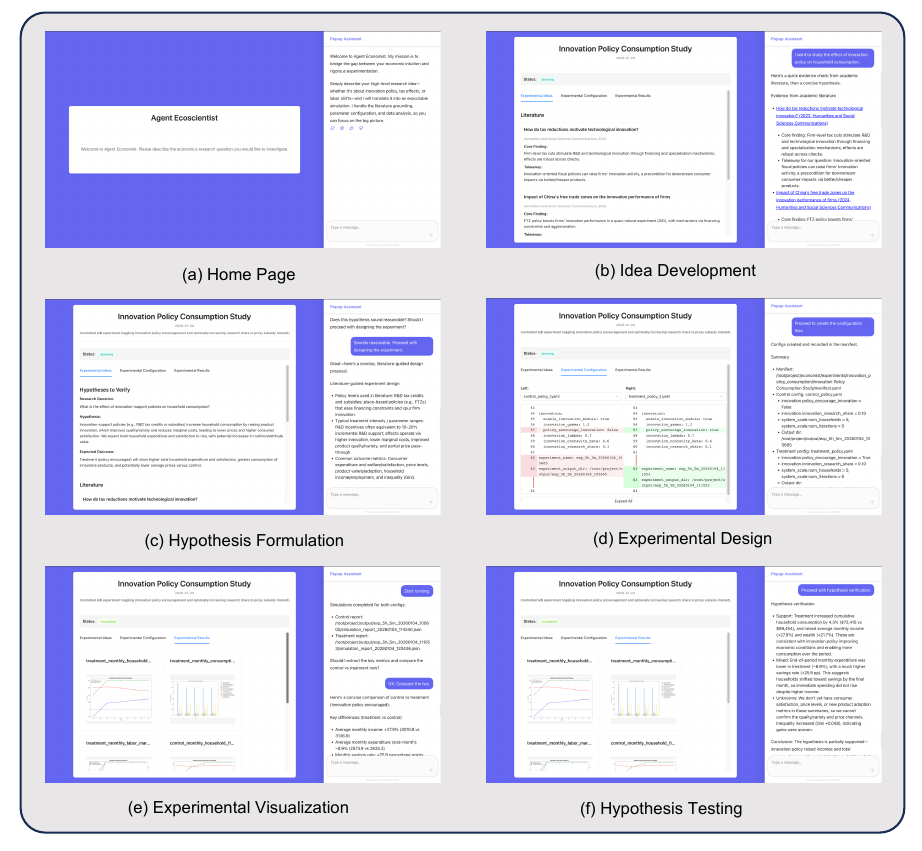}
    \caption{
The interactive research workflow of AgentEconomist, illustrating the end-to-end process from intuition input and ideation to experimental configuration, execution, visualization and hypothesis verification.}
    \label{fig:frontend}
\vspace{6mm}
\end{figure*}

\subsection{Human--Agent Interaction}
\label{sec:human_interaction}

Figure~\ref{fig:frontend_display} first presents the detailed page layout of the interface.
The right-side Copilot panel is used for human--agent dialogue, while the left-side research canvas provides a structured view of current experiment progress; workflow tabs allow users to switch across stages.
Within this layout, three core views support practical research operations:
\begin{itemize}[leftmargin=8pt]
    \item \textbf{Experimental Ideas.} This view organizes retrieved literature from the RAG pipeline together with the current research question, candidate hypotheses, and expected outcomes.
    \item \textbf{Experimental Configuration.} This view supports side-by-side parameter comparison across experiment groups (e.g., treatment vs. control), making group-level design differences explicit.
    \item \textbf{Experimental Results.} This view visualizes simulation outputs according to the selected metrics and presents corresponding numerical values in structured panels for inspection.
\end{itemize}

After clarifying the page-level components, Figure~\ref{fig:frontend} shows the end-to-end interaction workflow.
Users enter from the homepage, provide an initial idea for idea development, and confirm a reasonable hypothesis.
The system then conducts experimental design by constructing multiple experiment groups and assigning parameters to each group, followed by simulation execution and metric-based visualization.
Finally, the system performs hypothesis testing: if current evidence is insufficient, it proposes new directions for iterative refinement; if the hypothesis is adequately supported, the workflow is concluded and the task is marked as complete.
Throughout this process, the researcher remains in control, using the interface to guide, assess, and iterate on the inquiry.

\subsection{Computational Cost and Scalability}
\label{sec:cost_scalability}

We report computational cost along two dimensions: wall-clock runtime and token consumption.
For runtime, retrieval latency and LLM planning/design latency are relatively small in our implementation.
Retrieval relies on local vector similarity search (Qdrant with cosine), and these two stages are typically minute-level (around one minute each in common runs).

Simulation is the dominant contributor to end-to-end runtime.
Its duration is mainly determined by the number of households and the simulation horizon (months), and in our iterative workflow it also increases with the number of experimental iterations.

For usage cost, token consumption is dominated by the AgentEconomy simulation engine rather than by AgentEconomist's orchestration turns.
In our setting, AgentEconomist itself contributes relatively limited token overhead, while most tokens are consumed during simulation-time agent interactions inside AgentEconomy.
Under a representative setting (5 households and 5 experimental iterations), one full workflow takes about 20 minutes, and the reported 500K tokens primarily correspond to AgentEconomy-side simulation consumption.
In our current implementation, both runtime and total token usage scale approximately linearly with household count and iteration count, indicating predictable cost growth as workload increases.

\section{Experiments}
\label{sec:experiments}

We evaluate {AgentEconomist} as an end-to-end research copilot for economic simulation, focusing on whether it improves the \emph{intuition-to-experiment workflow}.
Our evaluation is guided by three research questions:
\begin{itemize}[leftmargin=8pt]
    \item \textbf{RQ1:} Can the system translate abstract economic intuitions into well-scoped, testable research ideas? 
    \item \textbf{RQ2:} Does the system effectively ground these hypotheses in relevant economic literature and coherent economic mechanisms? 
    \item \textbf{RQ3:} Can the system support end-to-end research workflows for user-posed questions, from intuition formulation and hypothesis refinement to executable experimentation and iterative analysis?
\end{itemize}

These questions cannot be adequately evaluated using static datasets or fixed benchmarks.
Economic research assistance is inherently interactive, open-ended, and path-dependent: the quality of an experiment depends on how intuitions are refined, grounded, and operationalized over multiple turns, rather than on a single correct output.
Accordingly, we adopt a mixed-methods evaluation design that combines controlled comparisons against strong LLM baselines with human-in-the-loop assessments based on real user interactions.

We first describe the experimental setup and data collection procedure (\S\ref{sec:exp_setup}), then report quantitative results on hypothesis quality across multiple economic dimensions (\S\ref{sec:translation_effect}), present a qualitative analysis of user feedback to characterize perceived trust, workflow support, and usability limitations (\S\ref{sec:qualitative}), and finally provide an end-to-end case study demonstrating that the system can surface interpretable scientific phenomena from user-posed intuitions (\S\ref{sec:case_study}).

\subsection{Experimental Setup}
\label{sec:exp_setup}

\textbf{Baselines.}
We evaluate {AgentEconomist} against strong general-purpose LLM baselines commonly used as standalone research assistants.
Participants were allowed to use the model they considered strongest for their own workflow (primarily {GPT-5.2} and {Gemini 3-Pro}).
These baselines are already highly capable and include strong retrieval and information-management abilities (e.g., web retrieval and long-context memory).
To reduce procedural bias, both baseline and AgentEconomist conditions were run under the same user-provided intuition and recorded with full interaction traces, following the paired protocol in Appendix~\S\ref{sec:prompt_demo}.

\textbf{Evaluation Metrics.}
For idea generation quality and grounding (RQ1/RQ2), we evaluate eight dimensions: \emph{Clarity \& Structure}, \emph{Literature Grounding}, \emph{Economic Logic}, \emph{Mechanism Completeness}, \emph{Hypothesis Specificity}, \emph{Novelty \& Insight}, \emph{Relevance \& Significance}, and \emph{Simulation Feasibility} (Appendix~\S\ref{sec:hypothesis_dimensions}).
For qualitative user experience analysis (RQ3), we use four dimensions: \emph{Perceived Advantages}, \emph{Trust and Credibility}, \emph{Pain Points and Limitations}, and \emph{Role Perception} (Appendix~\S\ref{sec:qualitative_protocol}).
All quality dimensions are scored with a 5-point Likert protocol (1=Very poor, 3=Acceptable, 5=Excellent); detailed scoring instructions are provided in Appendix~\S\ref{sec:scoring}. The user-side evaluation form is summarized in Appendix~\S\ref{sec:prompt_demo} (Figure~\ref{tab:prompt_questionnaire}).
Consistent with the appendix rubric, quantitative scoring is explicitly restricted to \emph{hypothesis-generation content only}, excluding downstream plots, policy discussion, and stylistic fluency.
For LLM-as-a-judge, we use the anonymous referee prompt in Appendix~\S\ref{sec:prompt_demo} (Figure~\ref{tab:prompt_referee}), including order-invariance, model-blindness, and a mandatory scope-compliance check.

\textbf{Participants.}
We recruited {15 participants} with backgrounds in economics, public policy, or related social science fields.
This sample size is consistent with prior expert-oriented ACL/CHI evaluations (typically around 15--20 participants) where tasks require domain expertise and extended interaction time~\citep{lu2025axis,hong2025game,pu2025ideasynth}.
Each participant interacted with the system by providing an economic intuition of their own choosing.
The study uses a paired within-subject design: each participant evaluates both conditions under matched scenarios, improving statistical power for direct comparison.
All participants completed a structured questionnaire; however, due to incomplete submissions, only {14 participants} provided full and usable interaction logs for hypothesis-level comparisons.
Following the appendix protocol, required records include baseline/AgentEconomist logs, the completed 8-dimension scoring table for both systems, and open-ended responses used for grounded-theory coding.

\subsection{Idea Quality and Grounding (RQ1 \& RQ2)}
\label{sec:translation_effect}

\begin{figure*}[!t]
    \centering
    \includegraphics[width=\textwidth]{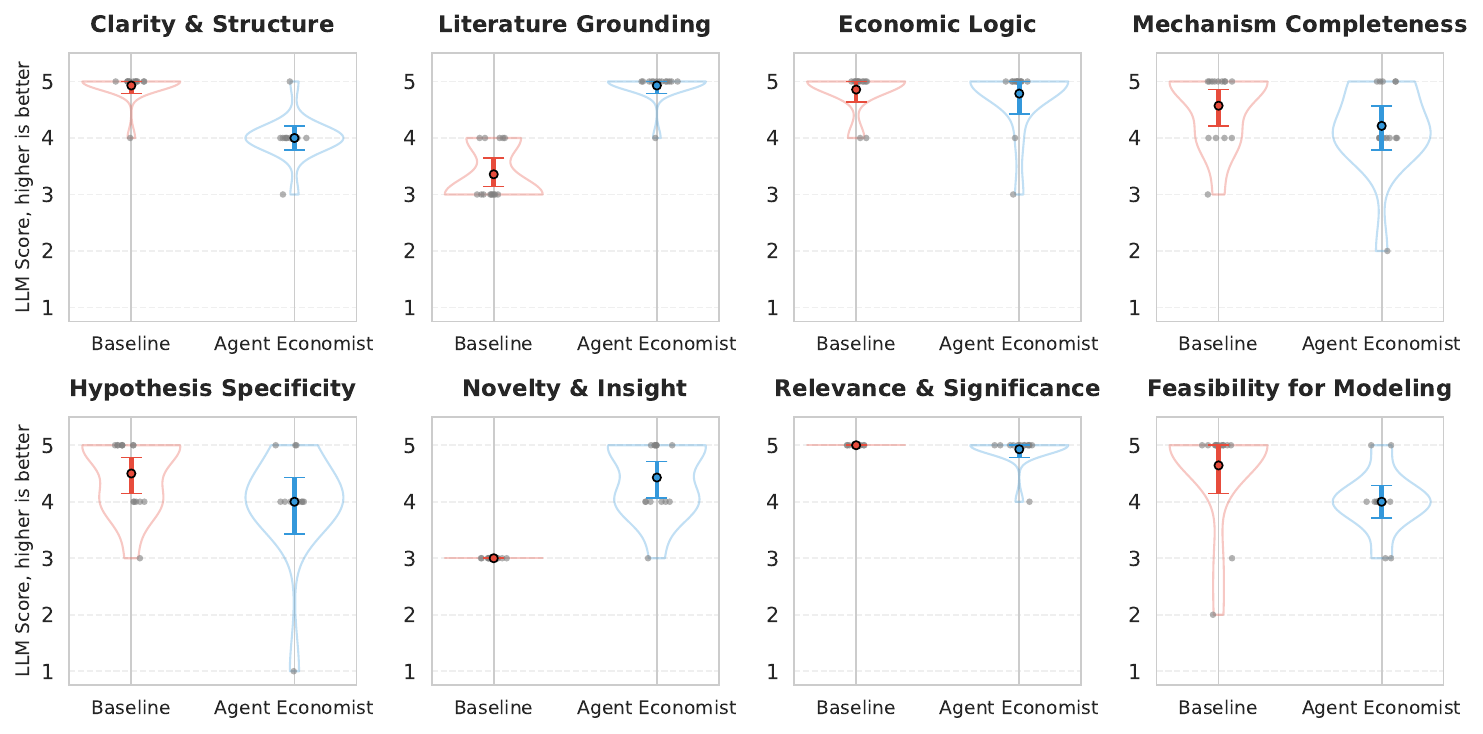}
    \vspace{2mm}
    \includegraphics[width=\textwidth]{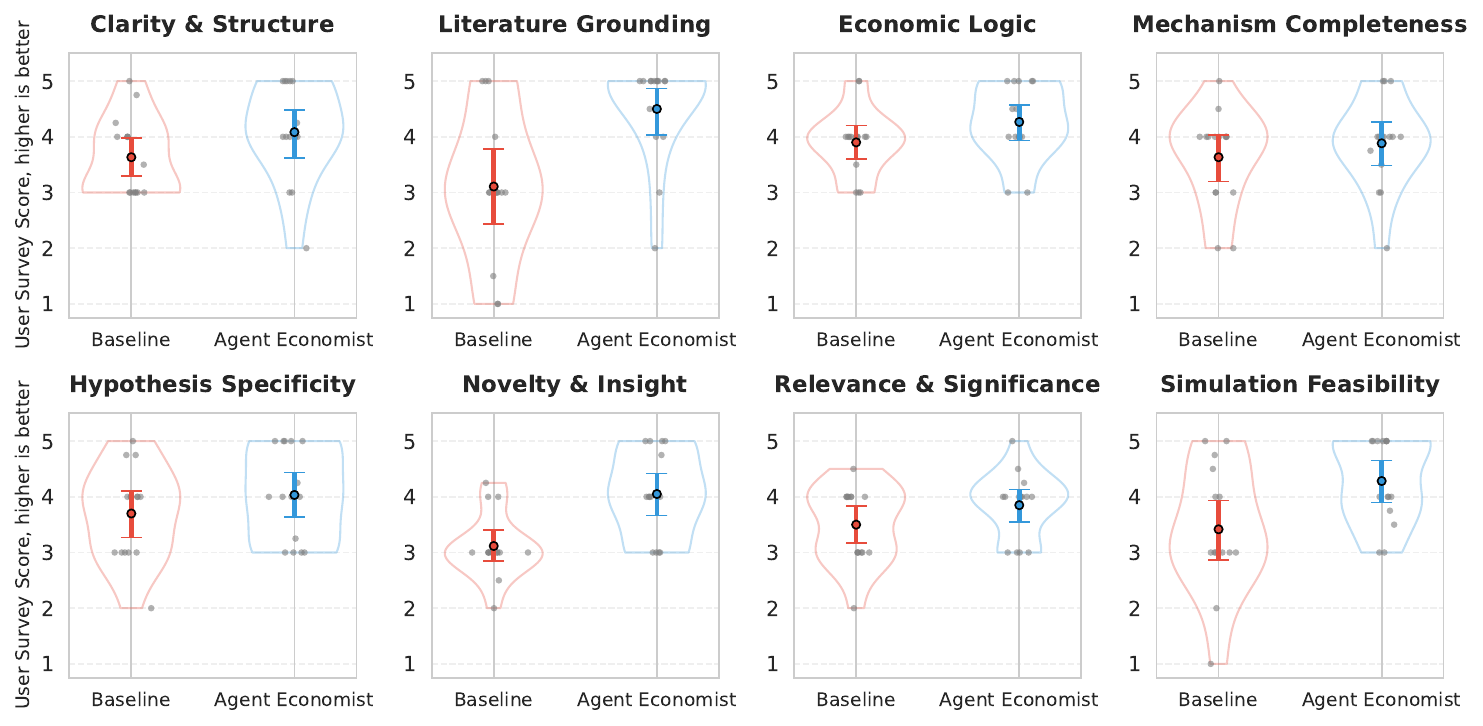}
    \vspace{-5mm}
    \caption{Distribution of hypothesis quality scores across eight evaluation dimensions.
    \textbf{Top}: LLM-based anonymous referee evaluation.
    \textbf{Bottom}: Human evaluation by study participants.
    Each subplot corresponds to one dimension.
    Across both judging protocols, AgentEconomist consistently outperforms the baseline, with the largest margins observed in \textit{Literature Grounding} and \textit{Novelty \& Insight}.}
    \label{fig:idea_quality}
\vspace{6mm}
\end{figure*}

This section evaluates whether AgentEconomist improves the quality of generated hypotheses and their grounding in economic literature.

\paragraph{Evaluation Protocol.}
To isolate hypothesis-level quality from downstream execution variability, we conduct controlled comparisons at the hypothesis generation stage.
Both AgentEconomist and baseline LLMs are prompted with identical user intuitions.
Generated hypotheses are evaluated along eight dimensions capturing economic rigor, insight, and implementability, using both an LLM-based anonymous referee and human judgments. LLM judging follows the referee template summarized in Section~\ref{sec:prompt_demo} (Figure~\ref{tab:prompt_referee}).
For statistical testing, we use a paired, two-sided Wilcoxon signed-rank test across dimensions, which is appropriate for our within-subject design and the ordinal nature of Likert-scale ratings.

\paragraph{Results.}
Figure~\ref{fig:idea_quality} summarizes the evaluation results.
To make the comparison explicit, we jointly report score improvements and statistical significance for the two primary dimensions below.

\textbf{Literature Grounding.}
This dimension measures whether cited theories and references are factual and contextually appropriate, whether they directly support the hypothesis, and whether hallucinated or weakly related citations are avoided.
Under LLM-based judging, the score improves from 3.36 to 4.93, with a significant difference ($p=0.00195<0.01$).
Under human evaluation, the score improves from 3.11 to 4.50, also with a significant difference ($p=0.0117<0.05$).
We attribute this large gain to three system-level mechanisms: (1) retrieval-augmented generation (RAG) over a curated economics corpus, which provides concrete and domain-relevant evidence before hypothesis drafting; (2) parameter-first hypothesis construction, which forces links between literature mechanisms and simulator-executable variables; and (3) structured memory, which preserves prior rationale and reduces drift across iterations.
Together, these constraints make hypotheses not only better cited, but also better anchored to the exact economic mechanism being tested.

\textbf{Novelty \& Insight.}
In our metric definition, this dimension evaluates whether a hypothesis goes beyond common claims and offers non-trivial, potentially counter-intuitive, or underexplored economic insight.
Under LLM-based judging, the score improves from 3.00 to 4.43, with a significant difference ($p=0.0039<0.01$).
Under human evaluation, the score improves from 3.12 to 4.05, with a significant difference ($p=0.0185<0.05$).
The improvement is consistent with AgentEconomist's interaction design: users begin from coarse intuitions, while the system expands candidate mechanisms through literature synthesis, then filters them with feasibility and identifiability constraints.
This workflow suppresses generic ``safe'' statements and encourages hypotheses that are simultaneously original and testable in simulation.
In contrast, baseline models more often produce fluent but conventional hypotheses, which are easier to write but less likely to expose genuinely new mechanism-level insights.

These results directly support our core objective: translating vague economic intuitions into hypotheses that are both literature-grounded and non-trivial, rather than merely improving rhetorical fluency.

\paragraph{Human--LLM Evaluation Differences.}
A systematic divergence arises in the Clarity \& Structure dimension.
While the LLM-based referee slightly favors the baseline, human evaluators rate AgentEconomist comparably or higher.
We attribute this difference to distinct evaluation priorities.
Baseline outputs tend to be shorter and rhetorically streamlined, whereas AgentEconomist produces denser hypotheses that explicitly articulate mechanisms, assumptions, and literature connections.
Although this additional structure may reduce surface-level clarity for automated judges, it aligns more closely with human researchers’ needs when developing grounded and novel economic hypotheses.

\subsection{End-to-End User Experience Analysis (RQ3)}
\label{sec:qualitative}

To examine how users perceive AgentEconomist as an end-to-end research framework, we conduct a formative qualitative analysis of open-ended questionnaire responses.
The questionnaire consists of five prompts:
{Q1} (Perceived Advantages),
{Q2} (Trust and Credibility),
{Q3} (Pain Points and Limitations),
{Q4} (Role Perception),
and {Q5} (Other Open Feedback). Methodological details for RQ3 are summarized in Appendix~\S\ref{sec:qualitative_protocol}

Not all participants answered every prompt (Q1: $n=8$, Q2--Q4: $n=7$, Q5: $n=3$).
Accordingly, our analysis focuses on recurring themes supported by multiple responses rather than estimating population-level prevalence.

\paragraph{Method.}
We apply an LLM-assisted grounded-theory workflow using the thematic-analysis template described in Section~\ref{sec:thematic} (Figure~\ref{tab:prompt_grounded}).
For each analysis round, we instantiate the template with aggregated, anonymized participant responses and require structured outputs including core themes, concise summaries, and supporting verbatim excerpts.
The coding procedure follows an iterative open-coding-to-grouping process: we first extract candidate concepts from response segments, then merge semantically related concepts into higher-level themes, and finally retain only themes supported by multiple participants.
To reduce over-interpretation, we treat qualitative outputs as hypothesis-generating evidence of user experience, not as objective performance measurements.
Accordingly, we report only patterns that are directly traceable to textual evidence and use quotes/excerpts as grounding anchors for each reported theme.
The thematic template specification is summarized in Appendix~\ref{sec:thematic}.

\paragraph{Findings.}
Participants consistently emphasized {grounded trust}, attributing increased confidence to literature-backed reasoning enabled by the RAG module.
A second dominant theme is {operationalization support}: users valued the system’s ability to translate vague intuitions into simulator-aligned experimental configurations.
Third, respondents highlighted {mechanistic scaffolding}, noting that AgentEconomist more explicitly articulated causal and behavioral chains than generic LLM outputs.
Together, these factors contributed to a perceived {role shift} from a conversational assistant toward a research assistant or collaborative expert spanning hypothesis formulation and experiment setup.

\paragraph{Pain Points and Implications.}
Feedback also revealed usability bottlenecks, including execution latency, limited process transparency, and occasional instruction-following drift over long interactions.
These observations suggest that future work should prioritize improving responsiveness, exposing interpretable intermediate states, and strengthening long-horizon intent preservation.

\subsection{Case Study (RQ3)}
\label{sec:case_study}

To strengthen empirical realism beyond architectural description, we provide a detailed end-to-end case study based on a real user session.
The researcher began with a high-level intuition: \emph{whether government support for innovation increases household consumption}.

AgentEconomist first grounded this intuition by retrieving and synthesizing relevant economic literature on innovation incentives, firm productivity, and income--consumption linkages.
Based on this context, the system formulated a concrete research hypothesis and automatically designed a controlled experiment by toggling a policy variable that enables government support for innovation, while keeping market structure fixed.
A treatment group with innovation support is compared against a control group without such support, and key outcome metrics are tracked over time.

When the innovation-support parameter is activated, the treatment condition exhibits higher cumulative household consumption (+4.3\%), substantially higher income (+27.9\%) and wealth (+21.7\%), a higher savings rate, and slightly increased inequality.

\begin{figure}[t]
    \centering
    \includegraphics[width=0.95\columnwidth]{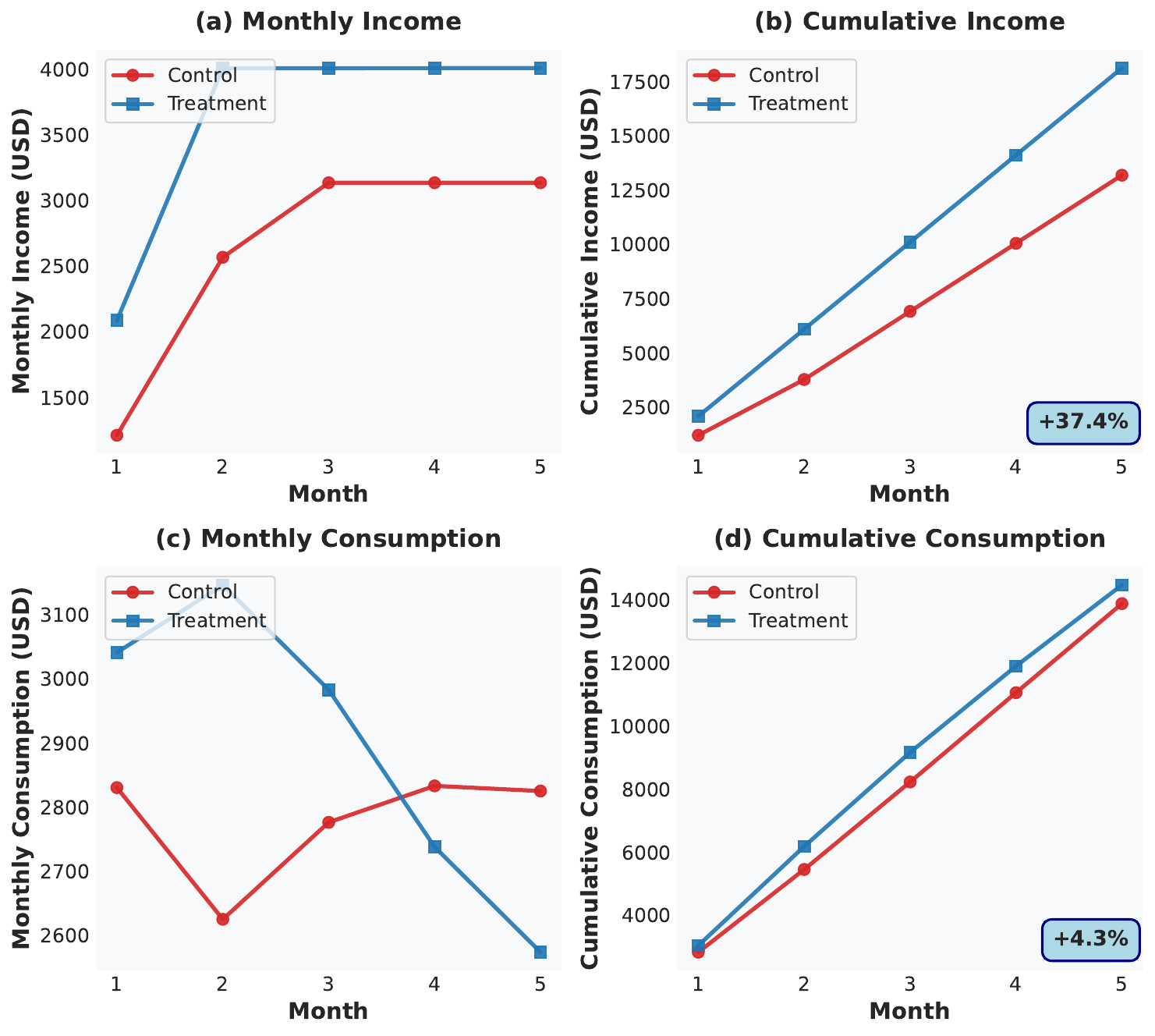}
    \caption{
    Income and consumption dynamics under innovation-support policies.}
    \label{fig:case_study_analysis}
\end{figure}

Figure~\ref{fig:case_study_analysis} shows the resulting income and consumption trajectories.
These outcomes are directionally consistent with established economic mechanisms: productivity-enhancing innovation policies increase firm output and income, which propagate to households through wage and profit channels, thereby raising consumption and wealth accumulation.
The slight increase in inequality is also theoretically plausible under innovation-driven growth, where gains may be distributed unevenly across agents.

Importantly, the purpose of this case study is not to claim that the simulator discovers new economic laws.
Rather, it demonstrates that (i) the system produces theoretically coherent and interpretable outcomes, (ii) effect magnitudes are explicit and quantifiable, and (iii) users retain control over parameterization so that assumptions can be inspected and adjusted.
In settings where theoretical constraints are critical (e.g., policy interventions with trade-offs or negative externalities), the system relies on literature grounding during hypothesis generation and supports iterative refinement of assumptions and parameters.

Overall, this case demonstrates that AgentEconomist enables an interpretable end-to-end research workflow—from intuition grounding and experimental design to simulation and mechanism-level analysis.

\section{Conclusion}

We presented AgentEconomist, an end-to-end agentic system that supports economic research by translating abstract intuitions into executable computational experiments.
The system adopts a human-in-the-loop design that decomposes the intuition-to-experiment workflow into literature-grounded idea development, experiment formalization, and execution, implemented by specialized agents.
Empirical evaluation shows that AgentEconomist improves hypothesis quality on the dimensions most critical to economic inquiry, particularly literature grounding and novelty of insight, while remaining effective in interactive research settings.
Throughout this process, the system remains controllable through executions and iterations explicitly mediated by the user, mitigating potential risks associated with autonomous behavior.
More broadly, this work highlights the importance of epistemic scaffolding in scientific AI systems, where supporting theory-grounded reasoning and iterative sense-making is often more valuable than end-to-end automation.
Our findings also suggest a practical design implication for scientific copilots: domain grounding and executable-constraint awareness should be treated as first-class components, not optional post-processing steps.
By explicitly coupling hypothesis generation with simulation feasibility checks and traceable iteration records, AgentEconomist helps users move from plausible ideas to verifiable experimental artifacts with less hidden trial-and-error.
We therefore view this system not as a replacement for researchers, but as an infrastructure layer for faster and more transparent scientific exploration in economics.
We hope this perspective informs future work on domain-grounded, interactive agents for scientific discovery.

\section*{Limitations}

This work has several limitations. 
First, our evaluation focuses on the intuition-to-experiment workflow in an agent-based economic simulation, and does not assess performance on real-world policy deployment or empirical data analysis. 
Second, while the user study provides evidence of the system’s usefulness in interactive research settings, the number of participants is limited and may not capture the full diversity of economic research practices.
Third, the effectiveness of AgentEconomist depends on the coverage and quality of the underlying literature corpus and simulation environment; domains or research questions that fall outside these resources may be less well supported.
Finally, although the system maintains structured memory across interactions, long-horizon alignment and execution efficiency remain constrained by current LLM capabilities and system latency.
Future work may address these limitations by expanding empirical evaluation settings, incorporating larger and more diverse user populations, and improving system robustness and scalability.



\bibliographystyle{ACM-Reference-Format}
\bibliography{main}

@article{lu2024ai,
  title={The ai scientist: Towards fully automated open-ended scientific discovery},
  author={Lu, Chris and Lu, Cong and Lange, Robert Tjarko and Foerster, Jakob and Clune, Jeff and Ha, David},
  journal={arXiv preprint arXiv:2408.06292},
  year={2024}
}

@article{gottweis2025towards,
  title={Towards an AI co-scientist},
  author={Gottweis, Juraj and Weng, Wei-Hung and Daryin, Alexander and Tu, Tao and Palepu, Anil and Sirkovic, Petar and Myaskovsky, Artiom and Weissenberger, Felix and Rong, Keran and Tanno, Ryutaro and others},
  journal={arXiv preprint arXiv:2502.18864},
  year={2025}
}

@article{swanson2025virtual,
  title={The Virtual Lab of AI agents designs new SARS-CoV-2 nanobodies},
  author={Swanson, Kyle and Wu, Wesley and Bulaong, Nash L and Pak, John E and Zou, James},
  journal={Nature},
  volume={646},
  number={8085},
  pages={716--723},
  year={2025},
  publisher={Nature Publishing Group UK London}
}

@article{axtell2025agent,
  title={Agent-based modeling in economics and finance: Past, present, and future},
  author={Axtell, Robert L and Farmer, J Doyne},
  journal={Journal of Economic Literature},
  volume={63},
  number={1},
  pages={197--287},
  year={2025},
  publisher={American Economic Association 2014 Broadway, Suite 305, Nashville, TN 37203-2425}
}

@article{farmer2009economy,
  title={The economy needs agent-based modelling},
  author={Farmer, J Doyne and Foley, Duncan},
  journal={Nature},
  volume={460},
  number={7256},
  pages={685--686},
  year={2009},
  publisher={Nature Publishing Group UK London}
}

@article{zeng2025mirrormind,
  title={MirrorMind: Empowering OmniScientist with the Expert Perspectives and Collective Knowledge of Human Scientists},
  author={Zeng, Qingbin and Fan, Bingbing and Chen, Zhiyu and Ren, Sijian and Zhou, Zhilun and Zhang, Xuhua and Zhen, Yuanyi and Xu, Fengli and Li, Yong and Liu, Tie-Yan},
  journal={arXiv preprint arXiv:2511.16997},
  year={2025}
}

@article{shao2025omniscientist,
  title={OmniScientist: Toward a Co-evolving Ecosystem of Human and AI Scientists},
  author={Shao, Chenyang and Huang, Dehao and Li, Yu and Zhao, Keyu and Lin, Weiquan and Zhang, Yining and Zeng, Qingbin and Chen, Zhiyu and Li, Tianxing and Huang, Yifei and others},
  journal={arXiv preprint arXiv:2511.16931},
  year={2025}
}

@inproceedings{zeng2025reviewrl,
  title={ReviewRL: Towards Automated Scientific Review with RL},
  author={Zeng, Sihang and Tian, Kai and Zhang, Kaiyan and Wang, Yuru and Gao, Junqi and Liu, Runze and Yang, Sa and Li, Jingxuan and Long, Xinwei and Ma, Jiaheng and others},
  booktitle={Proceedings of the 2025 Conference on Empirical Methods in Natural Language Processing},
  pages={16942--16954},
  year={2025}
}

@article{gao2025reviewagents,
  title={Reviewagents: Bridging the gap between human and ai-generated paper reviews},
  author={Gao, Xian and Ruan, Jiacheng and Zhang, Zongyun and Gao, Jingsheng and Liu, Ting and Fu, Yuzhuo},
  journal={arXiv preprint arXiv:2503.08506},
  year={2025}
}

@article{weng2024cycleresearcher,
  title={Cycleresearcher: Improving automated research via automated review},
  author={Weng, Yixuan and Zhu, Minjun and Bao, Guangsheng and Zhang, Hongbo and Wang, Jindong and Zhang, Yue and Yang, Linyi},
  journal={arXiv preprint arXiv:2411.00816},
  year={2024}
}

@article{zhu2025deepreview,
  title={Deepreview: Improving llm-based paper review with human-like deep thinking process},
  author={Zhu, Minjun and Weng, Yixuan and Yang, Linyi and Zhang, Yue},
  journal={arXiv preprint arXiv:2503.08569},
  year={2025}
}

@article{shao2025sciscigpt,
  title={SciSciGPT: advancing human--AI collaboration in the science of science},
  author={Shao, Erzhuo and Wang, Yifang and Qian, Yifan and Pan, Zhenyu and Liu, Han and Wang, Dashun},
  journal={Nature Computational Science},
  pages={1--15},
  year={2025},
  publisher={Nature Publishing Group US New York}
}

@inproceedings{guo2025ideabench,
  title={Ideabench: Benchmarking large language models for research idea generation},
  author={Guo, Sikun and Shariatmadari, Amir Hassan and Xiong, Guangzhi and Huang, Albert and Kim, Myles and Williams, Corey M and Bekiranov, Stefan and Zhang, Aidong},
  booktitle={Proceedings of the 31st ACM SIGKDD Conference on Knowledge Discovery and Data Mining V. 2},
  pages={5888--5899},
  year={2025}
}

@article{li2025datasetresearch,
  title={Datasetresearch: Benchmarking agent systems for demand-driven dataset discovery},
  author={Li, Keyu and Jiang, Mohan and Fu, Dayuan and Wu, Yunze and Hu, Xiangkun and Wang, Dequan and Liu, Pengfei},
  journal={arXiv preprint arXiv:2508.06960},
  year={2025}
}

@article{gao2025democratizing,
  title={Democratizing AI scientists using ToolUniverse},
  author={Gao, Shanghua and Zhu, Richard and Sui, Pengwei and Kong, Zhenglun and Aldogom, Sufian and Huang, Yepeng and Noori, Ayush and Shamji, Reza and Parvataneni, Krishna and Tsiligkaridis, Theodoros and others},
  journal={arXiv preprint arXiv:2509.23426},
  year={2025}
}

@inproceedings{su2025many,
  title={Many heads are better than one: Improved scientific idea generation by a llm-based multi-agent system},
  author={Su, Haoyang and Chen, Renqi and Tang, Shixiang and Yin, Zhenfei and Zheng, Xinzhe and Li, Jinzhe and Qi, Biqing and Wu, Qi and Li, Hui and Ouyang, Wanli and others},
  booktitle={Proceedings of the 63rd Annual Meeting of the Association for Computational Linguistics (Volume 1: Long Papers)},
  pages={28201--28240},
  year={2025}
}

@inproceedings{xu2025chatpd,
  title={ChatPD: An LLM-driven Paper-Dataset Networking System},
  author={Xu, Anjie and Ding, Ruiqing and Wang, Leye},
  booktitle={Proceedings of the 31st ACM SIGKDD Conference on Knowledge Discovery and Data Mining V. 2},
  pages={5106--5116},
  year={2025}
}

@article{weng2025deepscientist,
  title={Deepscientist: Advancing frontier-pushing scientific findings progressively},
  author={Weng, Yixuan and Zhu, Minjun and Xie, Qiujie and Sun, Qiyao and Lin, Zhen and Liu, Sifan and Zhang, Yue},
  journal={arXiv preprint arXiv:2509.26603},
  year={2025}
}

@article{hu2025survey,
  title={A survey of scientific large language models: From data foundations to agent frontiers},
  author={Hu, Ming and Ma, Chenglong and Li, Wei and Xu, Wanghan and Wu, Jiamin and Hu, Jucheng and Li, Tianbin and Zhuang, Guohang and Liu, Jiaqi and Lu, Yingzhou and others},
  journal={arXiv preprint arXiv:2508.21148},
  year={2025}
}

@article{huang2025deep,
  title={Deep research agents: A systematic examination and roadmap},
  author={Huang, Yuxuan and Chen, Yihang and Zhang, Haozheng and Li, Kang and Zhou, Huichi and Fang, Meng and Yang, Linyi and Li, Xiaoguang and Shang, Lifeng and Xu, Songcen and others},
  journal={arXiv preprint arXiv:2506.18096},
  year={2025}
}

@inproceedings{Singh2022SciRepEvalAM,
  title={SciRepEval: A Multi-Format Benchmark for Scientific Document Representations},
  author={Amanpreet Singh and Mike D'Arcy and Arman Cohan and Doug Downey and Sergey Feldman},
  booktitle={Conference on Empirical Methods in Natural Language Processing},
  year={2022}
}

@incollection{polanyi2009tacit,
  title={The tacit dimension},
  author={Polanyi, Michael},
  booktitle={Knowledge in organisations},
  pages={135--146},
  year={2009},
  publisher={Routledge}
}

@book{popper2005logic,
  title={The logic of scientific discovery},
  author={Popper, Karl},
  year={2005},
  publisher={Routledge}
}

@book{becker1976economic,
  title={The economic approach to human behavior},
  author={Becker, Gary S},
  volume={803},
  year={1976},
  publisher={University of Chicago press}
}

@book{north1990institutions,
  title={Institutions, institutional change and economic performance},
  author={North, Douglass C},
  year={1990},
  publisher={Cambridge university press}
}

@book{schelling2006micromotives,
  title={Micromotives and macrobehavior},
  author={Schelling, Thomas C},
  year={2006},
  publisher={WW Norton \& Company}
}

@article{samuelson1948foundations,
  title={Foundations of economic analysis},
  author={Samuelson, Paul Anthony},
  journal={Science and Society},
  volume={13},
  number={1},
  year={1948}
}

@article{harrison2004field,
  title={Field experiments},
  author={Harrison, Glenn W and List, John A},
  journal={Journal of Economic literature},
  volume={42},
  number={4},
  pages={1009--1055},
  year={2004},
  publisher={American Economic Association}
}

@book{duflo2011poor,
  title={Poor economics},
  author={Duflo, Esther and Banerjee, Abhijit},
  volume={619},
  year={2011},
  publisher={PublicAffairs New York}
}

@article{tesfatsion2006agent,
  title={Agent-based computational economics: A constructive approach to economic theory},
  author={Tesfatsion, Leigh},
  journal={Handbook of computational economics},
  volume={2},
  pages={831--880},
  year={2006},
  publisher={Elsevier}
}

@article{lebaron2006agent,
  title={Agent-based computational finance},
  author={LeBaron, Blake},
  journal={Handbook of computational economics},
  volume={2},
  pages={1187--1233},
  year={2006},
  publisher={Elsevier}
}

@inproceedings{tisue2004netlogo,
  title={Netlogo: A simple environment for modeling complexity},
  author={Tisue, Seth and Wilensky, Uri and others},
  booktitle={International conference on complex systems},
  volume={21},
  pages={16--21},
  year={2004},
  organization={Boston, MA}
}

@book{epstein1996growing,
  title={Growing artificial societies: social science from the bottom up},
  author={Epstein, Joshua M and Axtell, Robert},
  year={1996},
  publisher={Brookings Institution Press}
}

@book{railsback2019agent,
  title={Agent-based and individual-based modeling: a practical introduction},
  author={Railsback, Steven F and Grimm, Volker},
  year={2019},
  publisher={Princeton university press}
}

@article{chhikara2025mem0,
  title={Mem0: Building production-ready ai agents with scalable long-term memory},
  author={Chhikara, Prateek and Khant, Dev and Aryan, Saket and Singh, Taranjeet and Yadav, Deshraj},
  journal={arXiv preprint arXiv:2504.19413},
  year={2025}
}

@inproceedings{lu2025axis,
  title={Axis: Efficient human-agent-computer interaction with api-first llm-based agents},
  author={Lu, Junting and Zhang, Zhiyang and Yang, Fangkai and Zhang, Jue and Wang, Lu and Du, Chao and Lin, Qingwei and Rajmohan, Saravan and Zhang, Dongmei and Zhang, Qi},
  booktitle={Proceedings of the 63rd Annual Meeting of the Association for Computational Linguistics (Volume 1: Long Papers)},
  pages={7711--7743},
  year={2025}
}

@inproceedings{hong2025game,
  title={Game Development as Human-LLM Interaction},
  author={Hong, Jiale and Wu, Hongqiu and Zhao, Hai},
  booktitle={Proceedings of the 63rd Annual Meeting of the Association for Computational Linguistics (Volume 1: Long Papers)},
  pages={4333--4354},
  year={2025}
}

@inproceedings{pu2025ideasynth,
  title={Ideasynth: Iterative research idea development through evolving and composing idea facets with literature-grounded feedback},
  author={Pu, Kevin and Feng, KJ Kevin and Grossman, Tovi and Hope, Tom and Dalvi Mishra, Bhavana and Latzke, Matt and Bragg, Jonathan and Chang, Joseph Chee and Siangliulue, Pao},
  booktitle={Proceedings of the 2025 CHI Conference on Human Factors in Computing Systems},
  pages={1--31},
  year={2025}
}


\appendix
\section{Experiment Details}
\label{sec:a}

\subsection{Hypothesis Quality Dimensions}
\label{sec:hypothesis_dimensions}

Participants were asked to evaluate each generated hypothesis along the following eight dimensions.
For each dimension, participants were instructed to provide an independent score based on their own judgment, focusing on the hypothesis content rather than writing style or presentation.

\begin{itemize}[leftmargin=8pt]
\item \textbf{Clarity \& Structure:} Is the hypothesis logically organized and free of ambiguous or vague claims? \textit{High-score criterion:} professional language, explicit A$\rightarrow$B logic, and minimal rhetorical ambiguity.
\item \textbf{Literature Grounding:} Are referenced theories and studies real, correctly cited, and contextually appropriate (RAG grounding)? \textit{High-score criterion:} accurate citation usage, no hallucinated references, and clear evidential support for the hypothesis.
\item \textbf{Economic Logic/Soundness:} Does the proposed mechanism align with core economic principles (e.g., supply--demand consistency, rational behavior, budget constraints)? \textit{High-score criterion:} internally coherent economic logic that does not violate basic principles without explicit justification.
\item \textbf{Mechanism Completeness:} Does the hypothesis specify a complete transmission chain from treatment/intervention to outcomes? \textit{High-score criterion:} not only states outcome shifts, but also explains how micro-level agent behaviors generate macro-level effects.
\item \textbf{Hypothesis Specificity:} Is the hypothesis concrete and quantifiable rather than generic? \textit{High-score criterion:} specifies direction, magnitude, agents, or conditions (e.g., low-skill wages decrease by 5--10\% under substitution effects).
\item \textbf{Novelty \& Insight:} Does the idea provide non-trivial perspectives beyond well-known claims? \textit{High-score criterion:} introduces comparatively underexplored or counter-intuitive insights, potentially linked to contemporary contexts (e.g., AI, UBI).
\item \textbf{Relevance \& Significance:} Is the research question economically and policy relevant? \textit{High-score criterion:} addresses substantively important societal or policy problems rather than marginal scenarios.
\item \textbf{Simulation Feasibility:} Can the hypothesis be implemented within the current agent-based simulation framework? \textit{High-score criterion:} relies on programmable variables and available simulator parameters, without unobservable or non-implementable "magic" variables.
\end{itemize}

\subsection{Scoring Protocol}
\label{sec:scoring}

Each dimension is scored on a 5-point Likert scale:
1 = Very poor,
3 = Acceptable,
5 = Excellent.
Judges are instructed to base scores solely on the hypothesis-generation content and to ignore writing style, verbosity, or downstream analysis.

\subsection{Qualitative User Experience Analysis Protocol}
\label{sec:qualitative_protocol}

This qualitative evaluation is designed as a formative assessment to capture participants’ subjective experiences when interacting with AgentEconomist as a research framework.
Participants are asked to provide free-text responses based on their own interaction trajectories, without assuming complete task coverage or uniform interaction length.

\paragraph{Evaluation Dimensions.}
Open-ended questions are organized around four aspects of user experience:
\begin{itemize}[leftmargin=8pt]
    \item \textbf{Perceived Advantages:} Key differences compared to using a general-purpose LLM.
    \item \textbf{Trust and Credibility:} Factors influencing confidence in generated hypotheses and experimental designs.
    \item \textbf{Pain Points and Limitations:} Usability issues, missing features, or interaction difficulties.
    \item \textbf{Role Perception:} How users conceptualize the system (e.g., search engine, research assistant, collaborative expert).
\end{itemize}

\subsection{Participant Background and Compensation}
\label{sec:participant_background}

All participants in the user study were doctoral students actively working in the area of economic modeling or simulation-based economic research.
Each participant had at least six months of prior research experience related to economic simulation or computational economics, ensuring sufficient domain knowledge to meaningfully evaluate the system.

Participants were compensated for their time at a rate commensurate with their academic background and local standards.
We consider the payment to be adequate given the participants' demographic and level of expertise, and the study involved no deceptive practices or sensitive data collection.
All participants provided informed consent prior to participation and were explicitly informed about how their interaction data would be used for research and evaluation purposes.


\section{Prompt Processing and Templates}
\label{sec:prompt_demo}

To improve transparency and reproducibility, we report the exact templates used in our evaluation pipeline and specify how placeholders are instantiated.

\subsection{User Study Questionnaire Template}
To standardize user-side evaluation across systems, we provide a questionnaire template used in our study.
The full protocol combines baseline-vs-ours comparison, mixed quantitative/qualitative reporting, and open-ended reflections. The concrete template is shown in Figure~\ref{tab:prompt_questionnaire}.

\begin{figure*}[t]
\centering
\setlength{\fboxsep}{7pt}
\fcolorbox{black!35}{black!5}{%
\begin{minipage}{0.97\textwidth}
\footnotesize
\ttfamily
\textbf{User Study Questionnaire Template} \\
\textbf{Objective:} Compare AgentEconomist against a strong general-purpose LLM baseline \\
for end-to-end economic simulation research support (hypothesis generation, experiment design, execution, and analysis). \\

\textbf{Task Scenarios (choose one or self-define)} \\
(A) Policy intervention: UBI effects on labor supply and consumption. \\
(B) Structural change: technology progress, wage structure, and sectoral shifts. \\
(C) System calibration: monetary-policy intervention under abnormal inflation. \\

\textbf{Procedure} \\
1) Baseline phase (e.g., GPT-5.2 / equivalent): run the same task and record the full interaction. \\
2) AgentEconomist phase: run the same task and record retrieval outputs, hypothesis updates, \\
   configuration steps, execution feedback, and final summaries. \\

\textbf{[Critical Scope Restriction for Quantitative Scoring]} \\
Score \textbf{only hypothesis-generation content}. Do not score downstream simulation plots, \\
policy discussion, writing style, verbosity, or rhetorical confidence. \\

\textbf{Quantitative Scoring Form (1--5 Likert; 1=Very poor, 3=Acceptable, 5=Excellent)} \\
\vspace{1mm}
{\scriptsize
\renewcommand{\arraystretch}{1.15}
\begin{tabular}{p{0.20\textwidth} p{0.34\textwidth} p{0.28\textwidth} c c}
\hline
\textbf{Metric Name} & \textbf{Definition} & \textbf{Criteria for High Score} & \textbf{GPT} & \textbf{Economist} \\
\hline
Clarity \& Structure & Is the hypothesis logically organized and unambiguous? & Professional language; clear A$\rightarrow$B logic; minimal vague claims. &  &  \\
Literature Grounding & Are cited theories/references real and contextually appropriate? & Accurate citations, no hallucination, and references directly support the hypothesis. &  &  \\
Economic Logic/Soundness & Does the mechanism follow core economic principles (incentives, constraints, rationality)? & Economically coherent and does not violate common-sense logic without justification. &  &  \\
Mechanism Completeness & Is the treatment-to-outcome transmission mechanism explicitly described? & Explains how micro-level agent behavior produces macro-level outcomes. &  &  \\
Hypothesis Specificity & Is the hypothesis concrete and quantifiable rather than generic? & Includes specific direction/magnitude/conditions (e.g., low-skill wages decrease by 5--10\%). &  &  \\
Novelty \& Insight & Does the idea provide non-trivial or counter-intuitive insight? & Goes beyond common claims and introduces underexplored perspectives (e.g., AI/UBI-related mechanisms). &  &  \\
Relevance \& Significance & Is the question policy-relevant and practically important? & Targets economically meaningful issues rather than marginal topics. &  &  \\
Simulation Feasibility & Can the idea be implemented in the current agent-based simulation environment? & Uses implementable variables/parameters; no unprogrammable "magic variables." &  &  \\
\hline
\end{tabular}
}

\textbf{Qualitative Questions (open-ended)} \\
Q1 Core difference: compared with the baseline, what is AgentEconomist's main advantage in your workflow? \\
Q2 Trust and credibility: which system do you trust more for experiment design, and why? \\
Q3 Pain points and limitations: what difficulties or missing features did you encounter? \\
Q4 Role perception: does AgentEconomist feel like a search engine, junior assistant, or collaborative expert? Why? \\
Q5 Optional additional comments. \\

\textbf{Required Records} \\
- Baseline and AgentEconomist interaction logs (text/screenshots/exported PDF) \\
- Final quantitative table (8 dimensions, both systems) \\
- Qualitative responses (Q1--Q5)
\end{minipage}
}
\caption{Template of the user study questionnaire used for quantitative and qualitative evaluation, including scoring scope constraints and required reporting fields.}
\label{tab:prompt_questionnaire}
\end{figure*}

\subsection{LLM-as-a-Judge Template}
For hypothesis-quality scoring, we use an anonymous economics-referee template.
Its placeholders are filled with paired hypothesis-generation outputs under matched user intuitions.
Specifically, \{N\} and \{M\} denote the number of pages for System A and System B materials, and the content slots are populated with anonymized hypothesis excerpts produced by each system.
All instantiated prompts and retrieved materials are logged for traceability. The concrete template is shown in Figure~\ref{tab:prompt_referee}.

\begin{figure*}[t]
\centering
\setlength{\fboxsep}{7pt}
\fcolorbox{black!35}{black!5}{%
\begin{minipage}{0.97\textwidth}
\footnotesize
\ttfamily
\textbf{Role: Anonymous Economics Referee (LLM-as-a-Judge)} \\
You are acting as an anonymous, neutral, and rigorous economics referee. \\
Your task is to evaluate and compare the quality of \textbf{hypothesis-generation content} \\
produced by two systems in response to the same economic question. \\

\textbf{[Critical Scope Restriction]} \\
Evaluate \textbf{ONLY the generated hypotheses themselves}. Do NOT consider: \\
- downstream analysis, simulations, policy discussion, extensions, or follow-up content. \\

\textbf{[Evaluation Materials]} \\
- Materials are provided as images or PDF pages. \\
- First \{N\} pages belong to System A; following \{M\} pages belong to System B. \\
- Output order does NOT imply priority. Ignore system identity. \\

\textbf{[Evaluation Principles]} \\
1) Content-only evaluation: ignore style, verbosity, tone, or rhetorical confidence. \\
2) Order invariance: do not favor earlier or later materials. \\
3) Model blindness: do not infer system identity or sophistication. \\
4) Referee-level standards: judge as if reviewing an economics paper’s hypotheses. \\

\textbf{[Dimensions: score each 1--5 (integer)]} \\
(1) Clarity \& Structure \\
(2) Literature Grounding \& Factual Plausibility \\
(3) Economic Logic / Soundness \\
(4) Mechanism Completeness (at hypothesis level) \\
(5) Hypothesis Specificity \\
(6) Novelty \& Insight \\
(7) Relevance \& Significance \\
(8) Feasibility for Modeling or Simulation \\

\textbf{[Required Output Format: Markdown Only]} \\
- A dimension-wise score table comparing System A vs System B with brief justification. \\
- Overall assessment ($\leq$ 200 words). \\
- Bias \& scope compliance check: \\
  \quad a) evaluated hypotheses only? (Yes/No) \\
  \quad b) ignored identity and order? (Yes/No) \\
  \quad c) any dimensions hard to score due to insufficient detail?
\end{minipage}
}
\caption{Prompt template for LLM-based hypothesis-quality judging (anonymous economics referee).}
\label{tab:prompt_referee}
\end{figure*}

\subsection{Thematic Analysis Template}
\label{sec:thematic}
For qualitative analysis, we use a grounded-theory template to code participants' open-ended feedback.
Its placeholder is filled with aggregated and anonymized response text (e.g., participant-level excerpts such as P1, P2, ...), derived from post-study questionnaires and interviews.
The instantiated prompts are then used to produce structured theme summaries for downstream comparison and reporting. The concrete template is shown in Figure~\ref{tab:prompt_grounded}.

\begin{figure*}[t]
\centering
\setlength{\fboxsep}{7pt}
\fcolorbox{black!35}{black!5}{%
\begin{minipage}{0.97\textwidth}
\footnotesize
\ttfamily
\textbf{Role: Social Science Researcher (Grounded Theory)} \\
You are a social science researcher familiar with qualitative methods. \\
Please use grounded theory (Grounded Theory) to conduct a systematic thematic analysis \\
on the following interview materials. \\

\textbf{[Methodological Constraint]} \\
Remain highly sensitive to the data and avoid pre-set theoretical assumptions. \\
Ensure conclusions are grounded in the interview materials themselves. \\

\textbf{[Required Output Structure]} \\
- Core themes \\
- Theme summary \\
- Verbatim example sentences \\

\textbf{[Interview Materials]} \\
\{Aggregated anonymized responses, e.g., P1:, P2:, ...\}
\end{minipage}
}
\caption{Prompt template for LLM-assisted grounded-theory thematic analysis of participants' open-ended feedback.}
\label{tab:prompt_grounded}
\end{figure*}

\section{AI Assistance Disclosure}

AI-based tools were used in the preparation of this work for code development assistance and for grammatical and stylistic editing of the manuscript. All scientific content, experimental design, results, and conclusions were conceived, implemented, and verified by the authors.

\end{document}